\documentclass[floatfix,aps,prx,reprint,superscriptaddress,amsmath,amssymb]{revtex4-2}
\usepackage[T1]{fontenc}
\usepackage[utf8]{inputenc}
\usepackage{lmodern}
\usepackage[english]{babel}

\usepackage[hidelinks]{hyperref}
\usepackage[capitalize,nameinlink]{cleveref}

\usepackage{graphicx}
\graphicspath{{figs/}}
\usepackage[version=4]{mhchem}
\usepackage{tabularx}
\usepackage{siunitx}
\usepackage{bm}

\DeclareMathOperator{\sinc}{sinc}

\begin{document}

\title{Superconducting on-chip spectrometer for mesoscopic quantum systems}

\author{J. Griesmar}
\affiliation{$\Phi_{0}$, JEIP, USR 3573 CNRS, Collège de France, PSL University, 11, place Marcelin Berthelot, 75231 Paris Cedex 05, France}
\affiliation{Laboratoire de Physique de la Matière Condensée, CNRS, Ecole polytechnique, Institut Polytechnique de Paris, 91120 Palaiseau, France}
\author{R. H. Rodriguez}
\author{V. Benzoni}
\affiliation{$\Phi_{0}$, JEIP, USR 3573 CNRS, Collège de France, PSL University, 11, place Marcelin Berthelot, 75231 Paris Cedex 05, France}
\author{J.-D. Pillet}
\affiliation{Laboratoire de Physique de la Matière Condensée, CNRS, Ecole polytechnique, Institut Polytechnique de Paris, 91120 Palaiseau, France}
\author{J.-L. Smirr}
\affiliation{$\Phi_{0}$, JEIP, USR 3573 CNRS, Collège de France, PSL University, 11, place Marcelin Berthelot, 75231 Paris Cedex 05, France}
\author{F. Lafont}
\affiliation{Dept. of Condensed Matter Physics, Weizmann Institute of Science, Rehovot 7610001, Israel}
\author{Ç. Ö. Girit}
\affiliation{$\Phi_{0}$, JEIP, USR 3573 CNRS, Collège de France, PSL University, 11, place Marcelin Berthelot, 75231 Paris Cedex 05, France}
\email{caglar.girit@college-de-france.fr}



\date{\today}

\begin{abstract}
  Spectroscopy is a powerful tool to probe physical, chemical, and biological systems.
  Recent advances in microfabrication have introduced novel, intriguing mesoscopic quantum systems including superconductor-semiconductor hybrid devices and topologically non-trivial electric circuits.
  A sensitive, general purpose spectrometer to probe the energy levels of these systems is lacking.
  We propose an on-chip absorption spectrometer functioning well into the millimeter wave band which is based on a voltage-biased superconducting quantum interference device.
  We demonstrate the capabilities of the spectrometer by coupling it to a variety of superconducting systems, probing phenomena such as quasiparticle and plasma excitations. 
  We perform spectroscopy of a microscopic tunable non-linear resonator in the 40-50 GHz range and measure transitions to highly excited states.
  The Josephson junction spectrometer, with outstanding frequency range, sensitivity, and coupling strength will enable new experiments in linear and non-linear spectroscopy of novel mesoscopic systems.
\end{abstract}

\maketitle

\section{\label{intro}Introduction}
Soon after the theoretical prediction of the Josephson effect, researchers directly measured microwave emission from a superconducting tunnel junction~\cite{Yanson2} and demonstrated that such junctions can also be used as detectors of external radiation~\cite{GrimesRichardsShapiro}.
Silver and Zimmerman were the first to combine the emission and detection capabilities of Josephson junctions to make an integrated absorption spectrometer~\cite{SilverZimmerman}.
Using a voltage-biased superconducting point contact coated in powdered cobalt, they measured the nuclear magnetic resonance response of Co$^{59}$ at \SI{218}{\mega\Hz} in the point-contact current-voltage characteristic.
Deaver extended the technique to higher frequencies with a niobium point contact coated with a resonant absorber, measuring features in the current-voltage characteristic at \SI{0.85}{\milli\volt} corresponding to the absorber frequency \SI{416}{\GHz}~\cite{Deaver}.

\begin{figure}
\includegraphics[width=0.95\columnwidth]{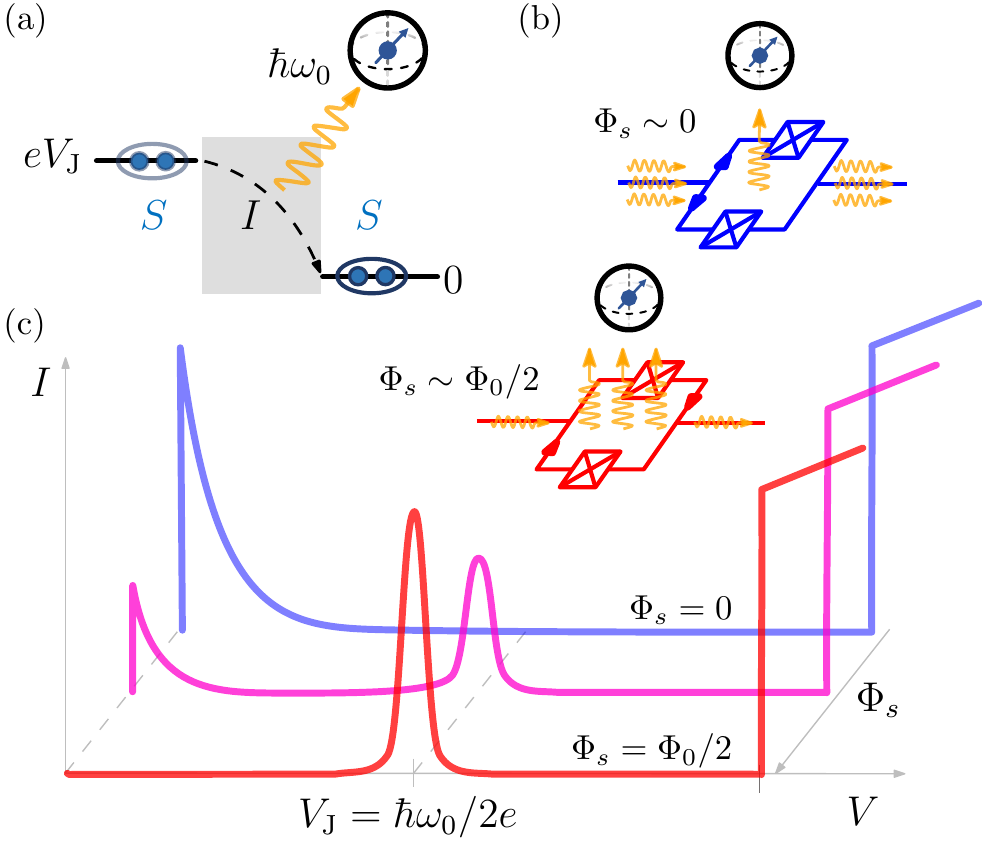}
\caption{\label{fig:principle}Principle of Josephson spectrometer.
  (a) Emission of photons with energy $2eV_{J}$ during Cooper pair tunneling in a Josephson tunnel junction (JJ) biased at voltage $V_{J}$.
  The device-under-test (DUT, Bloch sphere) can absorb photons when $2eV_J=\hbar\omega_{0}$.
  (b) A SQUID-based spectrometer (JJs, boxed crosses) excites electromagnetic modes which are predominantly in-plane for magnetic flux $\Phi_{s}\sim0$ (blue) and out-of-plane for $\Phi_{s}\sim\Phi_{0}/2$ (red).
  (c) For a DUT coupled to the out-of-plane mode the current-voltage characteristic of the Josephson spectrometer shows an increase in the height of the absorption peak at voltage $V_{J}=\hbar\omega_{0}/2e$ as the magnetic flux bias is adjusted from $\Phi_s=0$ to $\Phi_s=\Phi_0/2$, along with a reduction in the background signal at low voltages $V\sim0$.
}
\end{figure}

Despite these early efforts coupling Josephson junctions to bulk samples, there have been few applications of the technique for spectroscopy of mesoscopic systems, artificial quantum coherent structures.
Using specialized circuits in each case, researchers have measured Josephson plasma modes~\cite{Kleinsasser1}, resonances in microresonators~\cite{Edstam,Holst}, the energy levels of Cooper pair transistors~\cite{Lindell1,Billangeon2007b, Basset2}, spin-wave resonances~\cite{petkovic2009}, and Andreev bound states~\cite{Bretheau2013, vanWoerkom2017}.
Although Josephson junction based absorption spectroscopy has the potential to access a frequency range spanning \si{\GHz} to \si{\THz} with unprecedented sensitivity, progress has been impeded by several technical challenges.
The bandwidth is strongly limited by junction capacitance, parasitic electromagnetic modes are prevalent, and the low-frequency region is dominated by artifacts from the biasing circuit.
We overcome these difficulties, implementing a uniform coupling scheme, variable output power, and reduction of background resonances, in order to realize a general-purpose spectrometer for mesoscopic systems.

The Josephson relations~\cite{Josephson} imply that a superconducting tunnel junction biased at a DC voltage $V_{J}$ is comparable to an alternating current source of frequency $\omega_{J}=2eV_{J}/\hbar$ and amplitude $I_{0}$, the critical current.
The oscillating Josephson currents may excite resonances of the junction's electromagnetic environment, resulting in a change in the DC current $I_{J}$ which can be explained by energy conservation~\cite{Eck,Fiske,Yanson,Yanson2}.
The power dissipated in the junction's electromagnetic environment by the high-frequency currents must be balanced by the DC power $I_{J}V_{J}$ supplied by the source to maintain the voltage $V_{J}$~\cite{Werthamer1966,Werthamer1967,Likharev_Dynamics_1986}.

In a microscopic quantum view, the back and forth tunneling of Cooper pairs in a voltage-biased junction is accompanied by the emission and re-absorption of photons at an energy $\hbar\omega_{J}=2eV_{J}$~\cite{Ingold1994,BretheauPRB}.
The junction's electromagnetic environment may include a generic device-under-test (DUT) with a well-defined photon absorption peak, such as a two-level system with an energy difference $\hbar\omega_{0}$ between ground and excited states [\cref{fig:principle}(a)].
At resonance, $\omega_{J}=\omega_{0}$, the DUT can absorb photons emitted by the junction, with one Cooper pair tunneling per absorbed photon.
Relaxation from excited states of the DUT will result in a steady-state photon absorption rate $\Lambda$ which must be matched by a DC Cooper-pair tunneling current $I_{J} = 2e\Lambda$.
Because of the large voltage-to-frequency conversion ratio, $1/\Phi_0 = 2e/h \approx \SI{0.48}{\tera\Hz\per\milli\volt}$, where $\Phi_0$ is the magnetic flux quantum, the DC current-voltage $(I_{J},V_{J})$ characteristic of the Josephson junction (JJ) corresponds to the millimeter wave absorption spectrum of the DUT and its environment.

For a Josephson spectrometer based on a symmetric superconducting quantum interference device (SQUID)~\cite{Clarke}, as sketched in \cref{fig:principle}(b), consisting of two JJs (boxed crosses) in a superconducting loop, the direction of the emitted radiation will depend on the total static magnetic flux $\Phi_{s}$ threading the loop.
When the SQUID inductance is negligible, $\Phi_{s}$ is equal to the applied magnetic flux.
The phase difference $\delta_{s}$ between the two JJs in the loop is given by the total reduced flux, $\varphi_{s}=2\pi \Phi_{s}/\Phi_{0}$.
When the total flux is a multiple of the flux quantum, and there is non-zero bias voltage $V_{J}$, the junctions are in phase ($\delta_{s}=0\;\mathrm{mod}\;2\pi$) and an oscillating current of amplitude $2I_{0}$ and frequency $\omega_{J}$ flows in the biasing leads (common mode).
On the contrary when the SQUID is frustrated, with the total flux which is a multiple of half a flux-quantum, the junctions are out-of-phase ($\delta_{s}=\pi\;\mathrm{mod}\;2\pi$).
An oscillating current of amplitude $I_{0}$ circulates in the superconducting loop, generating an oscillating out-of-plane magnetic field (differential mode).

For a DUT coupled to the electromagnetic field above the spectrometer loop, only the differential mode can excite transitions.
At $\Phi_{s}\sim0$ few photons are coupled to the DUT but many propagate in the common mode along the biasing leads, \cref{fig:principle}(b) (blue schematic).
These photons may excite spurious electromagnetic modes in the biasing circuit, especially at lower frequencies where the impedance of the external environment may be large.
Therefore there will be significant DC current in the spectrometer current-voltage characteristic at low voltages and a negligible absorption peak at the DUT resonant voltage $V_{J}=\hbar\omega_{0}/2e$, as shown in the schematic \cref{fig:principle}(c) (blue, $\Phi_{s}=0$).

Near half a flux-quantum $\Phi_{s}\sim\Phi_{0}/2$ many photons couple to the DUT and few excite the common mode, \cref{fig:principle}(b) (red schematic), resulting in a current-voltage characteristic $(I_{J},V_{J},\Phi_{s})$ with a large DUT absorption peak and little background signal at low voltages, \cref{fig:principle}(c) (red, $\Phi_{s}=\Phi_{0}/2$).

The magnetic flux $\Phi_{s}$ simultaneously tunes the power delivered to the DUT by the spectrometer and decouples the spectrometer from the external, or ``off-chip,'' electromagnetic environment.
In the low power or linear spectroscopy regime, where $\Phi_{s}\sim0$, only the first few excited states of the DUT are probed.
For $\Phi_{s}\sim \Phi_0/2$ the power delivered to the DUT is large enough to access the non-linear spectroscopy regime and probe excited states.
Changes in the shape of an absorption peak with increasing power reveals the anharmonicities typically found in mesoscopic systems.


\begin{figure}
\includegraphics[width=\columnwidth]{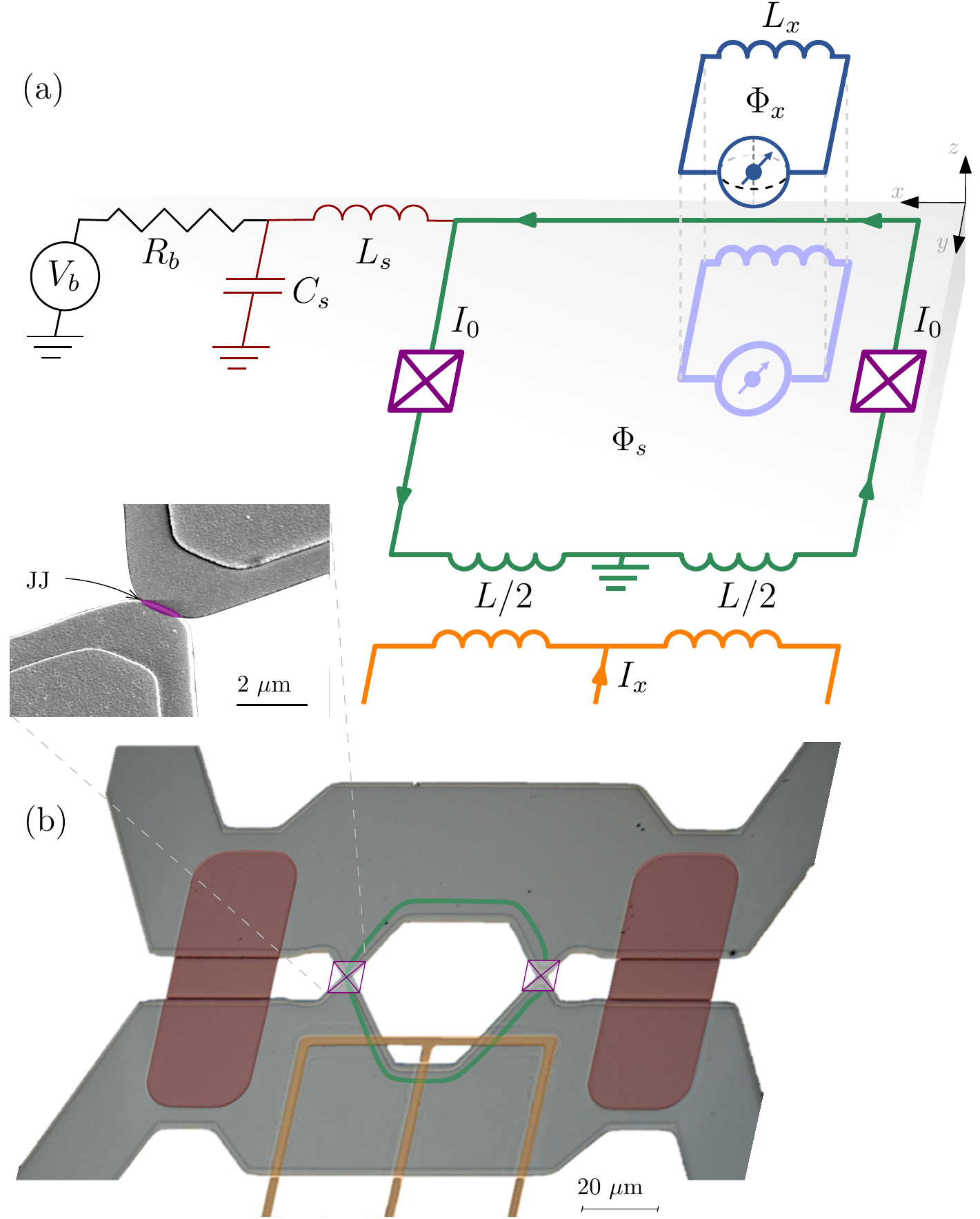}
\caption{\label{fig:implementation} Implementation of Josephson spectrometer.
  (a) Simplified electrical schematic.
  Oscillating currents generated by voltage-biased Josephson junctions (purple boxed crosses) flow in a superconducting loop (green) of inductance $L$ and couple inductively to an off-chip device-under-test (blue two-level system in superconducting loop of inductance $L_x$).
  The gradiometric flux line (orange, current $I_{x}$) in addition to an external coil allows independent control of the total magnetic flux $\Phi_s$ and $\Phi_x$.
  Filtering components ($C_s$ and $L_s$) reduce noise and spurious background signal.
  Bias circuit ($V_b$ and $R_b$) supplies DC current.
  (b) Colorized micrographs of spectrometer and junctions (inset).
  Aluminum superconductor is shown in gray, capacitors $C_{s}$ in red and gradiometric line in orange.
  The device-under-test (DUT) is not shown for clarity.
}
\end{figure}

\section{\label{impl}Spectrometer implementation}

\begin{figure}
\includegraphics[width=\columnwidth]{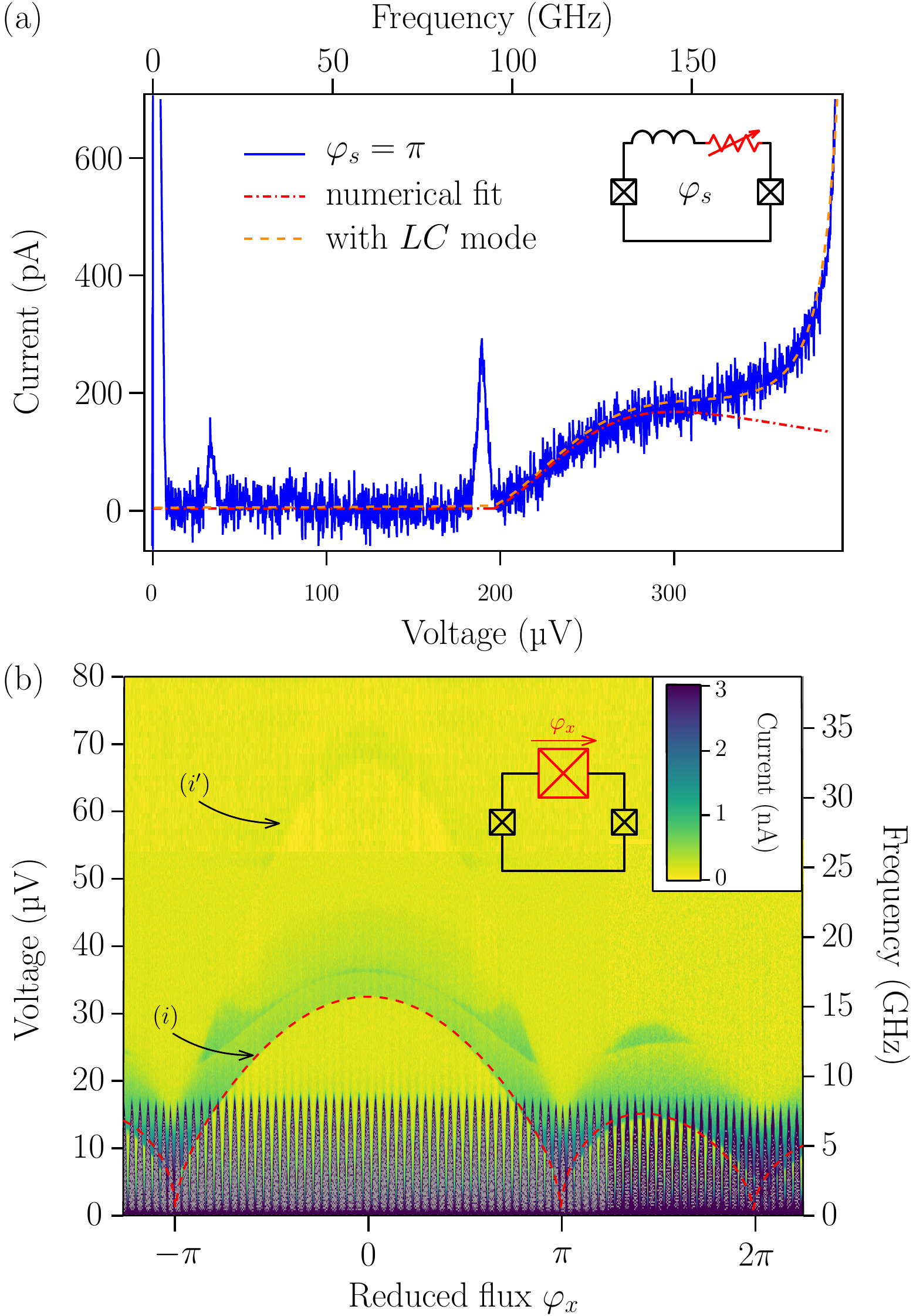}
\caption{\label{fig:spectro-meso}Josephson spectroscopy of mesoscopic systems.
  (a) Excitation spectrum of quasiparticles in the superconducting aluminum loop of Josephson spectrometer (inset) and numerical fit to Mattis-Bardeen theory and additional $LC$ resonator mode.
  In the schematic the dissipative Mattis-Bardeen conductance is represented by a red frequency dependent resistor in series with the loop inductance.
  (b) Spectral map showing first ($i$) and second harmonic ($i'$) of plasma resonance of a Josephson junction (red, inset) in spectrometer loop.}
\end{figure}

The advantages of using a frustrated SQUID as a Josephson spectrometer is that coupling to parasitic modes is reduced, the bandwidth is enhanced, and the power output is tunable.
Excitation of a general mesoscopic system can be implemented by coupling it to the spectrometer loop inductance via the generated magnetic field [\cref{fig:principle}], the magnetic flux [\cref{fig:implementation}], galvanically by inserting the DUT in the loop [\cref{fig:spectro-meso}], or capacitively with a parallel connection.
The coupling scheme can be adapted to the DUT, with low impedance mesoscopic devices such as superconducting ones suited for flux or galvanic coupling, whereas higher impedance devices, including spins or semiconductor crystals, better suited for capacitive coupling.

We describe in detail below a Josephson spectrometer with inductive coupling which allows for strong driving while allowing the DUT to be located on a separate chip.
A simplified electrical schematic for the Josephson spectrometer, including voltage and flux biasing circuits, is shown in \cref{fig:implementation}(a).
The DUT, indicated as a two-level system, is embedded in a superconducting loop of geometric inductance $L_{x}$ enclosing a total flux $\Phi_{x}$ and positioned above the spectrometer SQUID (purple boxed crosses in green superconducting loop) so as to couple only to the differential mode.
The total flux threading the DUT loop $\Phi_{x}$ will differ from the applied flux when screening of magnetic fields must be taken into account.

Due to the inductive coupling scheme the DUT does not need to be located on the same chip as the spectrometer, as shown in \cref{app:flipchip}, and the mutual coupling coefficient $\kappa$ will depend on the relative positions of the DUT and spectrometer [\cref{app:coupling}].
The maximum amplitude of the oscillating magnetic flux generated by the spectrometer is $I_{0}L$, where $L$ is the geometric inductance of the spectrometer loop.
For a DUT coupled to the spectrometer differential mode, the coupling energy can be approximated as $\kappa\sqrt{L/L_{x}}\, E_{J}\sin(\delta_{s}/2)\,\delta_{x}\cos \sigma$~\cite{Dmytruk}.
The Josephson energy of a spectrometer junction is $E_{J}=\varphi_{0}I_{0}$ and $\sigma$ is the average phase difference across the spectrometer.
The phase difference across the DUT, $\delta_{x}$, is equal to the total reduced flux $\varphi_{x}=2\pi\Phi_{x}/\Phi_{0}$.

An external coil (not-shown) combined with an on-chip gradiometric inductor (orange), with control current $I_{x}$, allows independent flux biasing of both spectrometer ($\Phi_{s}$) and DUT ($\Phi_{x}$).
The gradiometric line is designed to avoid inducing flux in the spectrometer loop but a calibration procedure is necessary to relate the control currents to the total flux in each loop, as well as to account for non-linear screening [\cref{app:phase-calibration}].
Passive filtering, both on-chip ($R_{b}, L_{s}, C_{s}$) and off-chip, help reduce spurious electromagnetic resonances in the spectrum.
The Josephson junctions comprising the spectrometer as well as the surrounding circuit, including filtering and bias, \cref{fig:implementation}(b), is fabricated using standard microlithography techniques [\cref{app:methods}].


\begin{figure}
\includegraphics[width=\columnwidth]{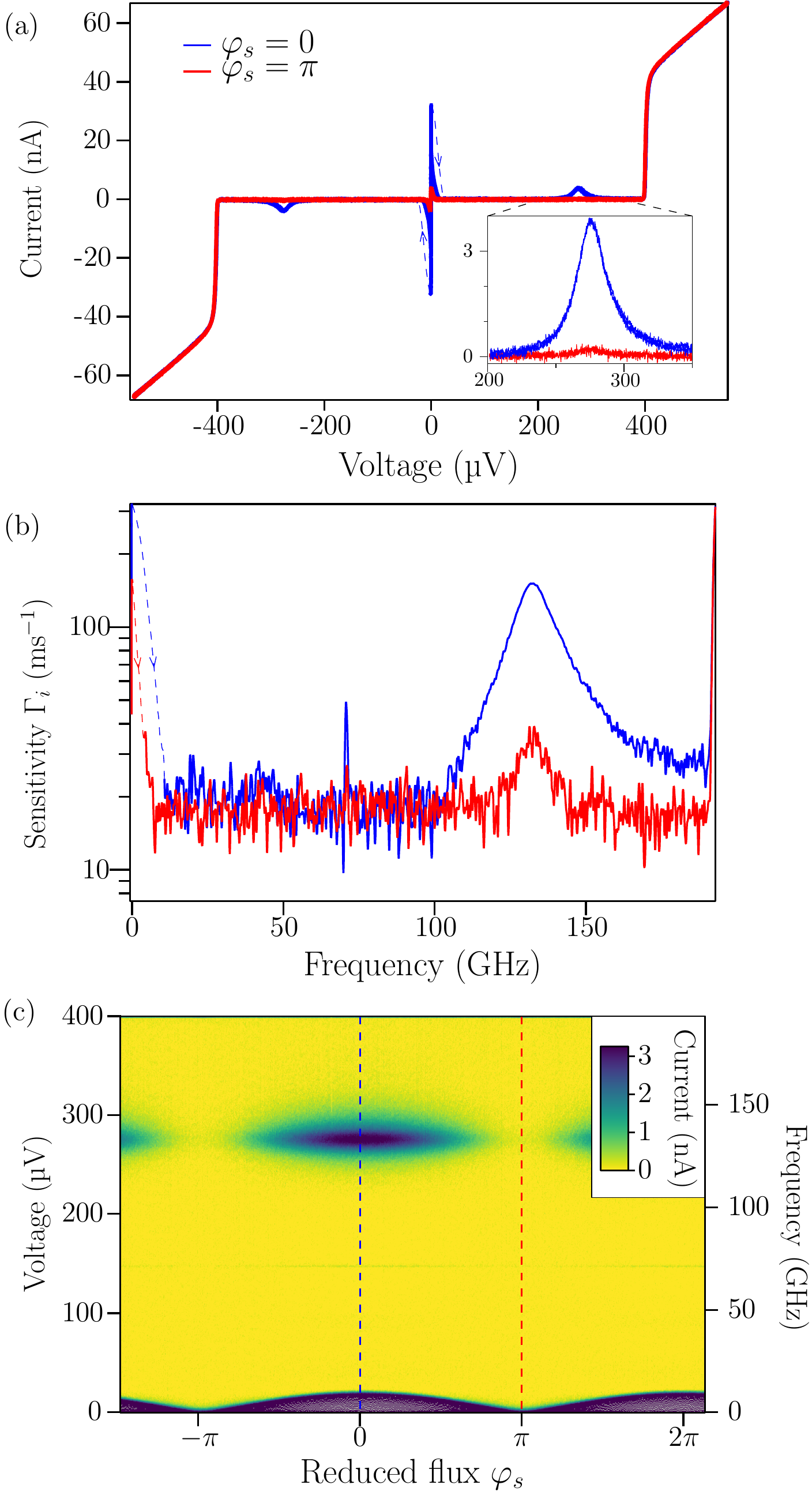}
\caption{\label{fig:bare-spectro}Characterization of Josephson spectrometer.
  (a) Measured current-voltage characteristic of the spectrometer at reduced flux bias $\varphi_s=0$ (blue) and $\varphi_s=\pi$ (red) in the absence of a DUT ($\varphi_s=2\pi\Phi_s/\Phi_{0}$).
  Inset shows a spurious series resonance which is smaller at $\varphi_{s}=\pi$.
  (b) Intrinsic sensitivity of spectrometer as a function of emission frequency, showing minimum theoretical detectable absorption rate of approximately \SI{20}{\per\ms} (\SI{1}{\Hz} measurement bandwidth).
  (c) Spectral map showing current as a function of bias voltage (emission frequency, right axis) and reduced flux.
  A spectrometer flux bias $\varphi_{s}=\pi$ avoids the resonance at \SI{132}{\GHz} and the switching region at low voltages (gray), providing a uniform background.
  The current-voltage characteristics of (a) are indicated by vertical dashed lines.
  Switching between the supercurrent branch and the sub-gap region is indicated with dashed lines in (a,b) and correspond to the gray lobes at low voltage where biasing is unstable in (c).
}
\end{figure}

We first demonstrate operation of a spectrometer in the absence of a DUT, \cref{fig:bare-spectro}.
Current-voltage characteristics are plotted for two spectrometer phases $\varphi_s=0,\pi$ in \cref{fig:bare-spectro}(a).
The phase-dependence of the supercurrent branch is as expected for a SQUID, \cref{fig:principle}(b), with a maximum at $\varphi_{s}=0$ and minimum at $\varphi_{s}=\pi$.
The non-zero switching current minimum is due to a small difference in the size of the two junctions comprising the SQUID.
In these and subsequent current-voltage characteristics, discontinuous switching (dashed lines) may occur at the top of peaks, including the supercurrent branch at zero voltage, due to the load line given by the bias resistor $R_b$ [\cref{fig:implementation}(b)].
Quasiparticle conduction results in a sharp increase in current at knee voltages $\pm 2\Delta/e \approx \pm\SI{400}{\micro\V}$, corresponding to a typical thin-film aluminum value for $\Delta$, the superconducting energy gap.

Near zero flux bias ($\varphi_s = 0$) and  $V_{J}=\SI{273}{\micro\volt}$, corresponding to \SI{132}{\GHz}, there is an extraneous absorption peak, highlighted in \cref{fig:bare-spectro}(a, inset), which can be identified as an electromagnetic resonance of the series biasing circuit.
Biasing at $\varphi_{s} = \pi$ greatly reduces the height of this peak and results in a virtually flat background over the whole frequency range, as shown in the log scale sensitivity plot, \cref{fig:bare-spectro}(b).

The intrinsic spectrometer sensitivity $\Gamma_{i}$ is defined as the minimum DUT absorption rate which exceeds the background noise in a given measurement bandwidth.
As the background noise increases it is more difficult to identify the current signal due to absorption by the DUT.
The \emph{intrinsic} current noise of a Josephson tunnel junction, similarly to a photodetector, is given by the shot noise spectral density $S_i=2eI_{bg}$, where $I_{bg}$ is the DC background current.
In order to improve sensitivity the dark count rate $I_{bg}/2e$ must be minimized, motivating a spectrometer design and fabrication process which results in small background currents.
For \cref{fig:bare-spectro}(b) the dark current of the spectrometer was measured with a long averaging time and converted to a shot-noise sensitivity in a $\nu=\SI{1}{\Hz}$ bandwidth.
The intrinsic spectrometer sensitivity away from the extraneous peak is approximately $\Gamma_{i}=\sqrt{S_i\nu}/2e\approx\SI{20}{\per\ms}$, corresponding to noise equivalent power (NEP) of \SI{1.3e-18}{\watt\per\sqrt\Hz} at~\SI{100}{\GHz}.
Sensitivity in the range \SIrange{100}{200}{\GHz} is better at phase bias $\varphi_{s}=\pi$ due to the reduced height of the spurious resonance.
The intrinsic sensitivity compares favorably to existing spectrometers [\cref{tab:spectro-comparison}], but the total sensitivity will be worse due to extrinsic factors such as current noise in the biasing circuit.
The external noise can be reduced by improved filtering, and the signal can be improved by cryogenic amplification and a photo-detector type current measurement circuit~\cite{Hobbs_2001,Curry2019}.

The full phase dependence of the spectrometer is shown in the spectral map \cref{fig:bare-spectro}(c), where the current, proportional to the absorption rate $\Lambda$, is plotted in color scale as a function of bias voltage (left axis) as well as frequency (right axis).
The current-voltage curves for $\varphi_s=0$ and $\pi$ are indicated by dashed lines.
Adjusting the spectrometer phase $\varphi_s$ not only changes the DC supercurrent, but also the power output of the high-frequency Josephson oscillations at non-zero voltage.
Since there is no DUT the spurious resonance at~\SI{132}{\GHz} (\SI{273}{\uV}) is the only remarkable feature.
The current peak due to absorption by this mode evolves \emph{in-phase} with the supercurrent peak at $V=0$, with a maximum at $\varphi_{s}=0$ and minimum at $\varphi_{s}=\pi$.
Such phase dependence for an absorption peak is a signature that the resonance couples to the common mode and is in series with the spectrometer.
At $\varphi_s=\pi$, the background is flat and the spectrometer is decoupled from this series resonance.

Unlike with previous realizations of Josephson spectrometers, the inductive nature of the coupling scheme used here ensures that the coupling strength at flux bias $\varphi_s=\pi$ is uniform and independent of frequency up to an intrinsic ``loop resonance'' \cite{ZimmermanSullivan,ZappeLandman,SongHurrell,TuckermanMagerlein}.
The loop mode is a lumped element resonance of the spectrometer circuit \cref{fig:implementation}(b) with angular frequency $1/\sqrt{LC_J/2}$ where $L$ is the loop inductance, proportional to the loop perimeter, and $C_J$ is the capacitance of a single junction.
The coupling to the loop mode is maximal at $\varphi_{s}=\pi$, resulting in an absorption peak which is 180° out of phase with the supercurrent peak and any series resonant modes.
Above the loop resonance frequency, the coupling to the DUT decreases rapidly as $1/\omega^2$.

For the spectrometer shown in~\cref{fig:bare-spectro} the loop mode frequency was designed to exceed $4\Delta/h\approx\SI{193}{\GHz}$ and so it is not visible in the map \cref{fig:bare-spectro}(c).
\cref{app:loop-mode} describes a spectrometer with a larger perimeter for which the loop mode is present in measurements.


\begin{figure*}
\includegraphics[width=\textwidth]{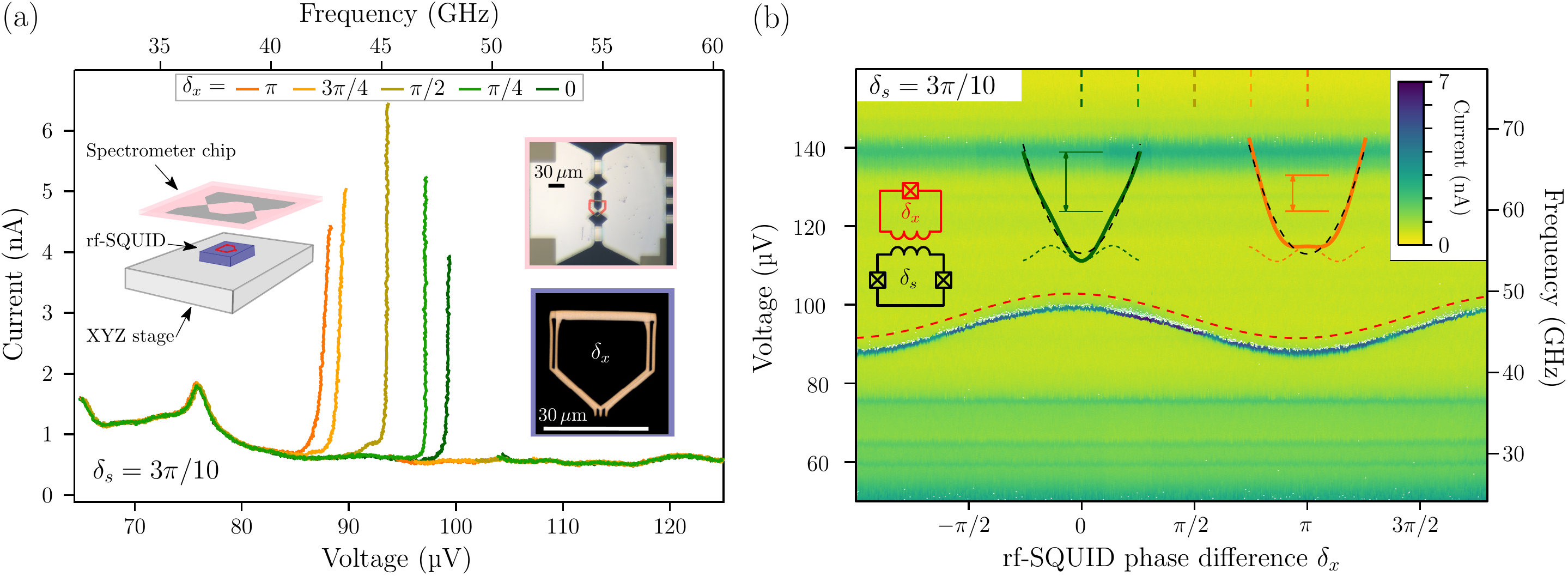}
\caption{\label{fig:linear-spectro}Linear spectroscopy of rf-SQUID.
  (a) Current-voltage characteristics of a Josephson spectrometer (top right) coupled to an rf-SQUID (bottom right) using a micro-positioner (schematic).
  The spectrometer is in the linear spectroscopy regime (low power, $\delta_{s}=3\pi/10$).
  The spectrometer phase difference $\delta_{s}$ and rf-SQUID phase difference $\delta_{x}$ are determined by calibrating the applied fields for cross-coupling and screening [\cref{app:phase-calibration}].
  Due to the switching instability, only half-peaks are resolved.
  (b) Spectral map of rf-SQUID (schematic, red) as a function of $\delta_x$ showing the periodic variation $\omega_{p}(\delta_x)$ and fit to model from text (dashed red line, offset by \SI{+3}{\micro\volt} for clarity).
  The model describes a circuit (left inset) for which the energy diagram (right inset) has transitions $h\omega_{p}$ (inset, energy diagram) which vary sinusoidally with $\delta_{x}$.
  Dashed vertical lines indicate spectra from (a).
}
\end{figure*}

To demonstrate how Josephson spectroscopy can be useful in probing mesoscopic systems, in \cref{fig:spectro-meso} we show the measured quasiparticle absorption spectrum of a short superconducting wire and the plasma resonance of a large Josephson junction.

In~\cref{fig:spectro-meso}(a), quasiparticles are excited in the superconducting loop of the spectrometer when it emits photons of energy $\hbar\omega_{J}$ greater than $2\Delta\approx h\cdot\SI{96}{\GHz}$, occurring for bias voltage $V = \hbar\omega_{J}/2e > \Delta/e\approx\SI{200}{\micro\volt}$.
This absorption is described by the Mattis-Bardeen theory for the frequency-dependent conductivity of superconductors with a BCS density of states~\cite{MattisBardeen}.
A numerical fit to the theory (dash-dotted line) closely follows the measured current rise at~\SI{200}{\micro\volt}.
The additional current relative to the Mattis-Bardeen prediction above~\SI{300}{\uV} is due to the spectrometer loop mode at \SI{199}{\GHz} and can be accounted for with an additional lumped-element $LC$ model (full fit, orange dashed line).
The fit parameters, including the normal state resistance (\SI{1.5}{\ohm}) and the loop mode frequency (\SI{199}{\GHz}), coincide with estimated values extracted from the junction capacitance (approximately \SI{40}{\femto\farad}) and loop geometry (approximate width \SI{5}{\um}, perimeter \SI{50}{\um}, inductance \SI{32}{\pico\henry})~\cite{Khapaev_2015}.
The two additional narrow peaks are attributed to inadvertent parasitic resonances~\cite{Griesmar_2018}.
For unconventional superconductors, similar Josephson spectroscopy measurements could determine the form of the quasiparticle density of states and reveal the symmetry of the order parameter~\cite{Sauls}.

In \cref{fig:spectro-meso}(b), another Josephson spectrometer (inset, black) is galvanically coupled to a large Josephson junction (inset, red) embedded in the spectrometer loop.
The area of the large junction is $\SI{12}{\um}\times\SI{500}{\nm}$ whereas each junction of the spectrometer has area $\SI{1}{\um}\times\SI{200}{\nm}$~\cite{Griesmar_2018}.
The spectral map shows Fraunhofer like absorption bands as a function of $\varphi_{x}$, the reduced flux which threads the large junction.
Although the applied flux is nominally oriented out-of-plane, a small fraction links the junction as stray in-plane flux due to misalignment and field distortion via the Meissner effect.
The Josephson plasma resonance, which disperses as $\omega_{p}(\varphi_{x})=\omega_{p0}\sqrt{\lvert\sinc \varphi_{x}/2\rvert}$, is identified as $(i)$ and fit with a plasma frequency $\omega_{p0}=2\pi\times\SI{16.2}{\GHz}$ (red dashed line), consistent with the estimated value,~\SI{15}{\GHz}.
A possible second harmonic $(i')$ of this mode as well as additional features in between $(i)$ and $(i')$ can be distinguished~\cite{Griesmar_2018}.
The narrow repetitive feature at low voltage corresponds to the flux modulation of the spectrometer critical current.
Previously, plasma resonances had only been measured with capacitive or series coupling schemes~\cite{Kleinsasser1,Bretheau2013,Lindell2}.

\begin{figure*}
\includegraphics[width=\textwidth]{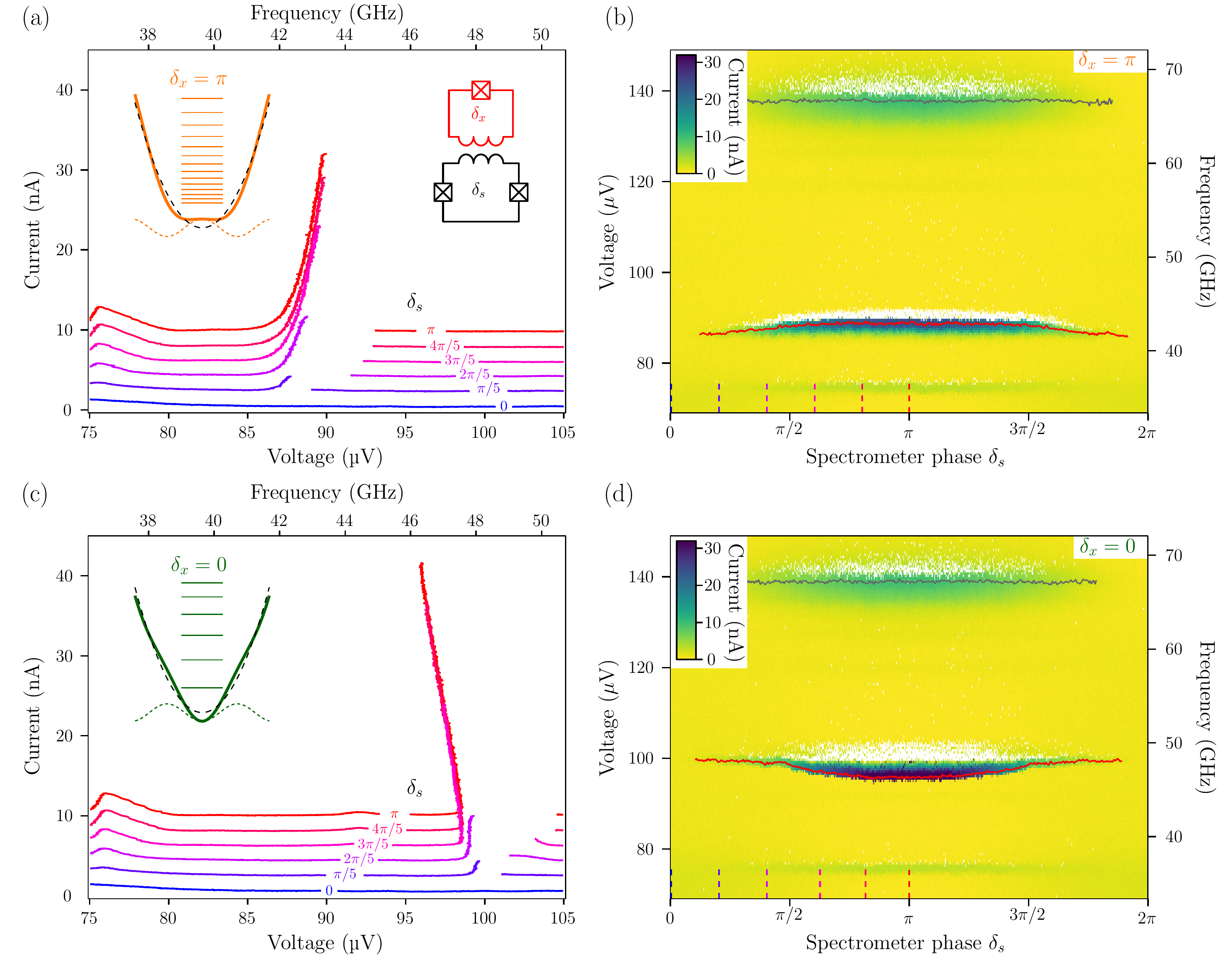}
\caption{\label{fig:nonlinear-spectro}Non-linear Josephson spectroscopy of rf-SQUID.
  (a) Spectra of rf-SQUID (inset, right) at $\delta_{x}=\pi$ as spectrometer power is increased, $\delta_{s}\rightarrow\pi$ (curves are offset by \SI{2}{\nA} for clarity).
  Peak maximum shifts to right due to increased level spacing of higher energy levels (inset, left).
  Due to the switching instability, only half-peaks are resolved.
  (b) Spectral map at $\delta_{x}=\pi$ showing position of rf-SQUID peak maximum as a function of $\delta_{s}$ (red line).
  For comparison the constant position of a spurious peak maximum (\SI{67.2}{\GHz}) for a linear resonance is also plotted (gray line).
  Unstable biasing regions are shown in white.
  (c) Spectra measured at rf-SQUID phase $\delta_{x}=0$.
  The peak maximum shifts to the left with increased power (back-bending) since the excited energy levels have reduced spacing (inset).
  (d) Corresponding spectral map ($\delta_{x}=0$) showing reduction in rf-SQUID absorption frequency at high power (red line) and reference peak (gray line).
}
\end{figure*}

\section{\label{rfSQUID}rf-SQUID Spectroscopy}

An rf-SQUID~\cite{Silver1965} is a prototypical quantum device which can be designed to have a tunable electromagnetic resonance frequency in the \SI{100}{\GHz} range~\cite{Clarke,Wendin_2007}.
This device consists of a single Josephson junction in a superconducting loop, making it ideal for coupling inductively to the Josephson spectrometer and testing its sensitivity.

\cref{fig:linear-spectro}(a) shows a schematic and images of a tunnel junction based rf-SQUID coupled to a Josephson spectrometer.
The rf-SQUID is fabricated on a separate chip from the spectrometer and carefully positioned near the spectrometer loop with an XYZ translation stage [\cref{app:flipchip}].
The mutual inductance will depend on the relative position between the rf-SQUID and spectrometer loop, as shown in~\cref{app:coupling}, and can be estimated from the geometry or determined by numerical simulation (3D-MLSI)~\cite{Khapaev_2001,Khapaev_2015}.

The critical current of the Josephson junction in the rf-SQUID is sufficiently large such that screening fields are non-negligible.
Therefore the total reduced flux threading the rf-SQUID, $\varphi_{x}$, as well as the reduced flux threading the spectrometer, $\varphi_{s}$, will depend in a non-linear fashion on the applied magnetic fields.
The corresponding phase differences across the rf-SQUID junction, $\delta_{x}$, and the two junctions of the spectrometer, $\delta_{s}$, are calibrated for independent control via currents applied to the coil and gradiometric flux line [\cref{app:phase-calibration}].
Although the total reduced fluxes $\varphi$ and corresponding phase differences $\delta$ can be used interchangeably, we reserve $\delta$ for graphs and text in which calibration has been performed and the total flux is not equal to applied flux.

In the linear spectroscopy regime $\delta_{s}$ is tuned away from the high-power point $\delta_{s}=\pi$ in order to weakly excite the rf-SQUID.
The absorption spectra in~\cref{fig:linear-spectro}(a) measured in this regime ($\delta_{s}=3\pi/10$) show a high-frequency resonance centered at approximately \SI{45}{\GHz}.
The resonance peak shifts to lower frequencies as the rf-SQUID phase difference $\delta_{x}$ is tuned away from $0$, giving a tuning bandwidth of~\SI{5}{\GHz}.
Due to the switching instability, only the left half of each peak is visible.

The rf-SQUID peak is distinguished from the background by its nearly sinusoidal dependence on $\delta_{x}$ as shown in the spectral map, \cref{fig:linear-spectro}(b).
Several small background peaks, independent of $\delta_{x}$, are present due to common-mode coupling to spurious microwave resonances in the gradiometric line and DC biasing lines when $\delta_{s}\approx 0$.

A schematic for the rf-SQUID and its coupling to the spectrometer, as well as the energy diagram, are shown in the insets of~\cref{fig:linear-spectro}(b).
The low-energy lumped element circuit model of an rf-SQUID consists of a geometric loop inductance $L_{x}$ in parallel with the flux-dependent non-linear Josephson inductance and the junction capacitance $C_{Jx}$~\cite{Clarke}.
The Josephson inductance is $L_{Jx}/\cos \delta_{x}$, where $L_{Jx}=\Phi_{0}/(2\pi I_{0x})$ and $I_{0x}$ is the junction critical current.
The full flux dependence of the resonance,
\begin{equation}
  \label{E1}
  \omega_x(\varphi_x)=\omega_{x0}\sqrt{1+\beta_{x}\cos \delta_{x}},
\end{equation}
where $\omega_{x0}=1/\sqrt{L_{x}C_{Jx}}$ and $\beta_x=L_x/L_{Jx}$,
is calculated by adding a small Josephson term to the harmonic oscillator potential.
When the contribution of the Josephson inductance to the total inductance is small, such that $\beta_x\ll1$, the central resonant frequency is approximately $\omega_{x0}$ and the tuning range is $\omega_{x0}(1\pm \beta_{x}/2)$.
The small size of the rf-SQUID loop, which determines $L_{x}$, and small junction area, proportional to $C_{Jx}$, leads to large $\omega_{x0}$ while $\beta_{x}$ determines the relative tuning bandwidth.

The energy dispersion in the data, \cref{fig:linear-spectro}(b), shows excellent agreement with the model, which is offset and indicated by a dashed line.
Fit details and parameters are given in~\cref{app:rfSQUID-spectrum}.
This measurement demonstrates the power of Josephson spectroscopy in coupling inductively to mesoscopic systems and measuring their high-frequency transitions.

The previous spectra were measured in the linear regime, at low spectrometer output power.
By adjusting the spectrometer phase to $\delta_{s}=\pi$ it is possible to strongly drive a non-linear DUT, inducing transitions to highly excited states.
Non-linear spectroscopy allows measuring power-dependent shifts in resonance frequency, or equivalently for quantum systems the change in level spacing with average occupation.
To enter this regime the coupling must be strong and the losses small enough such that the absorption rate at a given power is larger than the relaxation rate.

\cref{fig:nonlinear-spectro} shows absorption spectra of the rf-SQUID as a function of spectrometer output power.
In~\cref{fig:nonlinear-spectro}(a) and (b), the rf-SQUID phase difference is $\delta_{x} = \pi$, where the resonant frequency in the linear regime is at its minimum, approximately $\omega_{x0}(1-\beta_{x}/2)$.
The absorption peak for the lowest trace in~\cref{fig:nonlinear-spectro}(a) ($\delta_{s}=0$) is absent because no power is coupled to the rf-SQUID, as sketched in~\cref{fig:principle}(c) (blue).
At low power, $\delta_{s}=\pi/5$, the shape of the peak is identical to the one shown in the linear regime, \cref{fig:linear-spectro}(a) (orange, $\delta_{s}=3\pi/10$), except for a scaling factor.
As the spectrometer power increases ($\delta_{s}\rightarrow\pi$) the absorption peak bends to the right and the maximum peak position shifts towards higher frequency. 
We associate this shift with the excitation of higher energy states of the rf-SQUID, which due to anharmonicity, have larger level spacing when $\delta_{x}=\pi$ and for large level number (energy ladder, inset) [\cref{app:rfSQUID-spectrum}].
As shown in the inset the sum of the quadratic potential due to the inductance $L_{x}$ and the Josephson cosine potential of amplitude $E_{Jx}$ results in a decrease in convexity at the potential minimum.
Whereas low-lying states therefore have a smaller transition frequency, higher energy states do not ``see'' the bottom of the potential well and recover the energy spacing arising from the bare resonance, $\omega_{x0}=1/\sqrt{L_{x}C_{Jx}}$. 
The spectral map \cref{fig:nonlinear-spectro}(b), in which the peak maximum (red line) is overlaid on the rf-SQUID resonance, shows how the frequency shifts upward from the minimum as $\delta_{s}$ approaches $\pi$, even though the rf-SQUID phase difference $\delta_{x}=\pi$ is fixed.
For reference a peak corresponding to a spurious linear resonance at~\SI{67.2}{\GHz} (gray overlay) shows no variation in frequency and only its amplitude scales with spectrometer power.

On the contrary, for $\delta_x=0$ the rf-SQUID energy level are more closely spaced at higher level number [\cref{fig:nonlinear-spectro}(c), inset] due to the increase in convexity at the bottom of the potential well [\cref{app:rfSQUID-spectrum}].
This results in a backward bending of the rf-SQUID absorption peak in the current-voltage characteristics \cref{fig:nonlinear-spectro}(c) as spectrometer power is increased ($\delta_{s}\rightarrow\pi$).
This region of back-bending is generally not accessible with a conventional spectrometer, for which one expects discontinuities in the spectrum at points where the slope is vertical, such as at jumps of a Duffing oscillator.
Along a back-bending peak, the Josephson frequency decreases with increasing bias voltage $V_{b}$ [\cref{fig:implementation}(a)], which is possible because of the series bias resistor $R_{b}$.
This may allow probing ``hidden'' parts of certain non-linear resonances with the Josephson spectrometer.
In the spectral map \cref{fig:nonlinear-spectro}(d) for $\delta_{x}=0$ the back-bending results in the peak maximum moving downwards as power is increased (red line).

For a general non-linear DUT, analyzing the shape of a resonance peak as a function of spectrometer power provides information about relaxation rates in addition to the DUT excited state spectrum.
In the steady state the overall relaxation rate for transitions at frequency $\omega$ should be equal to the absorption rate $\Lambda$, otherwise the peak is not stable.
Further theoretical analysis allows fitting the measured spectra of the rf-SQUID at high power, quantifying the relaxation rates, and understanding the role of the bias resistor~\cite{Dmytruk}.

\section{\label{conc}Conclusion}
The sensitivity of the spectrometer, as well as several other figures of merit, are superior to those of conventional spectrometers (Table~1).
The low residual quasiparticle density in Josephson tunnel junctions leads to the high sensitivity and ultra low NEP.
Furthermore the Josephson spectrometer bandwidth spans \SI{1}{\GHz} to $2\times2\Delta/h$, corresponding to the onset of the quasiparticle branch in the current-voltage characteristic.
Although the Josephson oscillations only decay logarithmically at higher frequencies, sensitivity is reduced due to the large quasiparticle background current.
The soft upper frequency bound $4\Delta/h$ is approximately \SI{200}{\GHz} for aluminum and \SI{1.4}{\THz} for niobium.
The two principle limitations of the spectrometer are the necessity for low temperature and low magnetic field in order to maintain superconductivity.
By using alternative thin-film superconductors such as niobium, NbN~\cite{Kim_enhanced_2021}, or even YBCO~\cite{Cybart_nano_2015}, with higher critical field and transition temperature it would be possible to extend the operating range.

As for the Josephson spectrometer emission linewidth, an upper bound can be estimated from a current-voltage characteristic by measuring the linewidth of a small but narrow Lorentzian absorption peak.
This gives~\SI{100}{\MHz} which is consistent with thermal noise from the on-chip resistive lines combined with measurement noise.
A direct microwave measurement of the emission linewidth of a single Josephson junction with non-resistive leads gives a linewidth less than~\SI{10}{\MHz}, but the junction has stronger coupling to common-mode parasitic resonances.
The linewidth can be further reduced to less than~\SI{1}{\kHz} by injection locking to a conventional microwave source as with Shapiro steps~\cite{Shapiro}.

Additionally the spectrometer design allows strong coupling to microscopic mesoscopic systems without necessitating that the DUT be fabricated or located on the same chip.
As a result the spectrometer can also be reused with different DUTs.
We have demonstrated spatial sensitivity of the Josephson spectrometer by displacing the DUT [\cref{app:coupling}] to change the coupling strength, opening the way to spatially-resolved microwave spectroscopy via modifications to existing scanning SQUID implementations~\cite{Kirtley_2016,Anahory_2020}.

Whereas we demonstrated the capabilities of the spectrometer by coupling it to a microscopic oscillator of extremely high frequency and tuning range, novel physics can be probed in other mesoscopic systems.
These include hybrid superconductor-semiconductor based topologically non-trivial circuits~\cite{Krogstrup_2015}, multi-terminal superconducting weak links~\cite{riwar_multi-terminal_2016,Pillet2019}, Josephson Weyl circuits~\cite{Peyruchat2020,Fatemi2020}, and superconductors such as \ce{UPt3} or \ce{Sr2RuO4} which are posited to exhibit unconventional pairing~\cite{Kallin_2016,Sauls}.
A plethora of non-superconducting systems can also be probed with the Josephson spectrometer: high-frequency nano-mechanical and piezo-electric resonators~\cite{Oconnell_quantum_2010}, two-level systems in dielectrics~\cite{Lisenfeld_observation_2015}, NV centers~\cite{Kubo_2010}, spin-wave resonances~\cite{petkovic2009}, magnons~\cite{Tabuchi_quantum_2016}, and other localized or collective excitations in semiconductors.
Measurement of the excitation spectra of these systems may answer important open questions.
Josephson spectroscopy in general will lead to the exploration and detection of novel elementary excitations and quasiparticles.

\begin{table*}
  \caption{\label{tab:spectro-comparison}Comparison of Josephson Junction Spectrometer to Conventional Systems. Adapted from Hubers~\cite{Hubers2011}.}
  \begin{ruledtabular}
    \begin{tabular}{rp{2cm}p{2cm}p{2cm}p{2cm}p{2cm}}
      Spectrometer\footnotemark[1] & Frequency coverage (\si{\tera\Hz}) & Linewidth (\si{\mega\Hz}) & Maximum Brightness\footnotemark[2] (\si{\milli\watt/\mega\Hz}) & Minimum NEP\footnotemark[3] (\si{\watt}/$\sqrt{\si{\Hz}}$) & Minimum Detectable Absorption\footnotemark[4] \\
      \hline
      CW BWO & $0.1-1.5$\footnotemark[5] & $< 1$ & 20 & $10^{-13}$ & $10^{-9}$ \\
      CW multiplier & $0.1-2.6$\footnotemark[5] & $< 1$ & 20 & $10^{-13}$ & $10^{-8}$ \\
      CW QCL & $1-5$\footnotemark[5] & $< 0.1$ & 100 & $10^{-13}$ & $10^{-9}$ \\
      CW Photomixer & $0.1-2$ & $> 1$ & $10^{-3}$ & $10^{-14}$ & $10^{-6}$ \\
      p-type Ge\footnotemark[6] & $1-4$ & $< 1$ & $1000$ & $10^{-13}$ & $10^{-8}$ \\
      TDS\footnotemark[6] & $0.1-5$ & $3000$ & $10^{-19}$ & $10^{-16}$ & $10^{-7}$ \\
      FTS & $0.3-20$ & $3000$ & $10^{-14}$ & $10^{-13}$ & $10^{-4}$ \\
      \textbf{Josephson}\footnotemark[7] & $\mathbf{0.001-1.4}$ & $\mathbf{\ll 1}$ & $\mathbf{10^{-1}}$ & $\mathbf{10^{-19}}$ & $\mathbf{10^{-11}}$ \\
  \end{tabular}
\end{ruledtabular}
\footnotetext[1]{Abbreviations: CW, continuous wave; BWO, backward wave oscillator; QCL, quantum cascade laser; TDS, time domain spectrometer; FTS, Fourier transform spectrometer.}
\footnotetext[2]{The source output power divided by the linewidth.}
\footnotetext[3]{Noise Equivalent Power (NEP): the required spectral density of detected power to match the background noise power density (lower is better).}
\footnotetext[4]{Ratio of minimum detectable absorbed power over source output power. Value averaged over frequency interval \SIrange{0.1}{2}{\tera\Hz}.}
\footnotetext[5]{A single instrument cannot cover the entire frequency range.}
\footnotetext[6]{Pulsed source.}
\footnotetext[7]{Assuming niobium tunnel junctions and injection locking to reduce linewidth. With aluminum tunnel junctions the maximum frequency is \SI{0.2}{\THz} and without locking, the linewidth is less than \SI{100}{\MHz}.}
\end{table*}

\begin{acknowledgments}
  We thank Victor Brar, Landry Bretheau, Olesia Dmytruk, Daniel Estève, Marcelo Goffman, Fréderic Pierre, Fabien Portier, Hugues Pothier, and Marco Schiro for critical feedback.
  This project has received funding from the European Research Council (ERC) under the European Union's Horizon 2020 research and innovation programme (grant agreement 636744).
  The research was also supported by IDEX Grant No. ANR-10-IDEX-0001-02 PSL, a Paris ``Programme Emergence(s)'' Grant, the Office of Naval Research under Award No. N00014-20-1-2356, and the Thomas Jefferson Fund, a program of FACE Foundation launched in collaboration with the French Embassy.
  F.~L. acknowledges funding from the People Programme (Marie Skłodowska-Curie Actions) of the European Union’s Seventh Framework Programme (FP7/2007-2013) under REA grant agreement PCOFUND-GA-2013-609102, through the PRESTIGE programme coordinated by Campus France.
  We acknowledge the SEM Service of ESPCI.
\end{acknowledgments}




\appendix


\section{\label{app:methods}Methods}

The spectrometer and rf-SQUID were patterned with a combination of direct write laser microlithography and electron-beam lithography (Magellan 400 Thermofisher, SEM Service, ESPCI, Paris).
On-chip resistive leads (HfPd), capacitor and junction dielectrics (aluminum oxide), and superconductors (Al) were deposited in an electron beam evaporator.
Samples were cooled in closed-cycle dilution refrigerators with base temperature between \SIrange{10}{50}{\milli\K}.
Noise was filtered with a combination of custom cabling, shielding, and sample holders incorporating lumped-element and distributed filters.
Broadband silicon capacitors (\SI{100}{\nano\farad}, Murata) are wirebonded directly to the spectrometers in order to shunt high frequency noise and reduce the Josephson emission linewidth.
Current-voltage characteristics were measured with low-noise voltage and transimpedance amplifiers in a differential configuration when possible.

\begin{figure}
  \centering
\includegraphics[width=0.75\columnwidth]{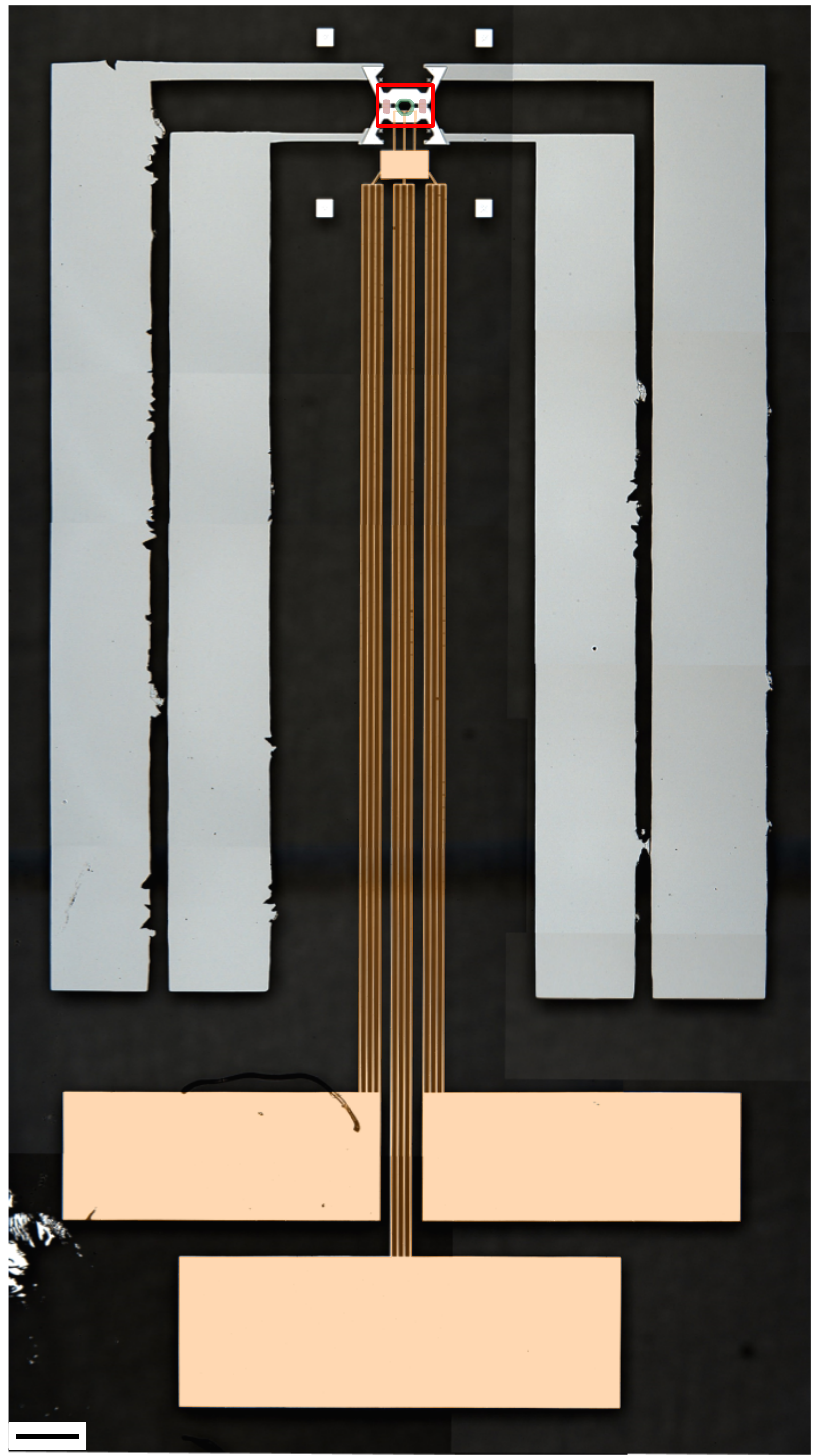}
\caption{\label{fig:spectro-photo}Photomicrograph of spectrometer.
  The aluminum spectrometer, in the central boxed region corresponding to \cref{fig:implementation}(a), is connected to an external circuit via hafnium-palladium resistive leads (gray).
  Current to the gradiometric flux coil is supplied via aluminum leads (colored light orange).
  Scale bar is \SI{200}{\um}.
}
\end{figure}

On-chip $LRC$ filters to tailor the high-frequency electromagnetic environment are implemented as shown in \cref{fig:implementation} and \cref{fig:spectro-photo}.
These filters are designed to reduce electromagnetic resonances coupling to the common mode while allowing DC biasing.
Close to the spectrometer SQUID, short inductances $L_{s}\approx\SI{30}{\pico\henry}$ reduce shunting the differential mode Josephson oscillations while capacitors $C_{s}\approx\SI{0.5}{\pico\farad}$ shunt the common-mode oscillations [\cref{fig:implementation}].
Beyond these two lumped elements a distributed low-impedance lossy transmission line made of HfPd with DC resistance \SI{200}{\ohm} dampens any common-mode oscillations leaking off chip [\cref{fig:spectro-photo}].
The Hf(45 nm)/Pd(25 nm) stack is not superconducting at base temperature and has a sheet resistivity of $\SI{4}{\ohm}/\square$.
Further details are provided in~\cite{Griesmar_2018}.

\section{\label{app:flipchip}Flip-chip assembly}

\begin{figure}
\centering
\includegraphics[width=\columnwidth]{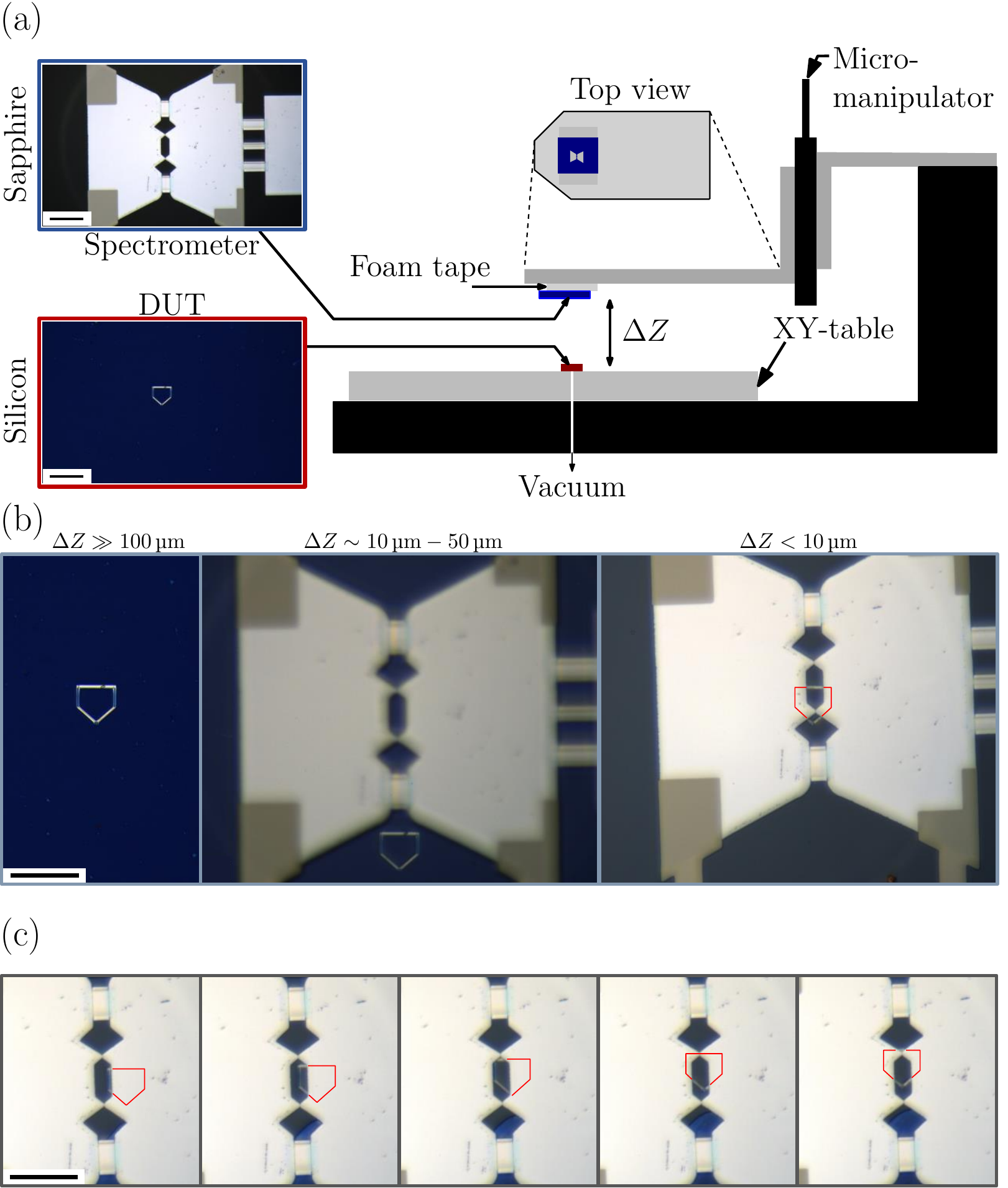}
\caption{\label{fig:flipchip}Flip-chip assembly.
  (a) Schematic of setup.
  (b) Reducing the spectrometer-DUT separation during alignment.
  (c) Fine positioning of the DUT.
  Scale bars correspond to~\SI{50}{\um}.
}
\end{figure}

The flip-chip setup to attach a device under test (DUT) to the spectrometer is shown in~\cref{fig:flipchip}.
The assembly consists of an optical microscope (not shown), the DUT, the spectrometer, and a micromanipulator~[\cref{fig:flipchip}(a)].
The DUT, here an rf-SQUID on a silicon chip, is placed facing upwards on the microscope manual XY translation table and held in place by a vacuum system.
The Josephson spectrometer, fabricated on a transparent sapphire chip, is mounted around the opening of a rigid support fixed on a manual micrometric vertical stage.
The vertical stage controls the separation $\Delta Z$ between both chips and the XY table allows positioning the DUT over the central loop of the spectrometer.
The opening, or window, where the spectrometer is attached (\emph{Top View}) allows monitoring the alignment under the microscope field of vision.

Optical images of the alignment procedure are shown in~\cref{fig:flipchip}(b).
As the chip separation $\Delta Z$ is decreased, the spectrometer enters into the field of view and is brought to focus for a separation within the depth of field.
The middle panel shows the DUT offset from the spectrometer for clarity.

\cref{fig:flipchip}(c) shows how the position of the DUT relative to the spectrometer can be adjusted within a few micrometers (minimum position change in this sequence is about \SI{4}{\um}).
For a small DUT the relative position determines the strength of the magnetic flux or magnetic field coupled to the DUT from the spectrometer [\cref{app:coupling}].

Once alignment is complete and the DUT chip is in contact with the spectrometer chip, a small drop of resist is deposited at the edge of the chips with a fine wire.
The droplet spreads via capillary action and glues the chips together upon drying.
At this point vacuum is disabled and the spectrometer chip, with the DUT glued in place, is removed from the tape on the rigid support.
The size and orientation of the DUT chip must be such that it does not cover the wirebond pads on the spectrometer chip [\cref{fig:spectro-photo}].
The ensemble is now ready for mounting in the sample box and wirebonding.

This setup is designed for aligning the spectrometer to a small, thin DUT which can be placed on a flat chip, and is suitable for standard superconducting or semiconducting devices.
Large samples, such as \si{\mm} or \si{cm} scale crystals or compounds, can be placed directly over the spectrometer chip and glued in place without alignment.
DUTs in solution can be dispersed directly or spin-coated onto the spectrometer chip.
The flip-chip assembly can also be modified to individually place a DUT on the spectrometer.
The rigid support is outfitted with a probe tip to which a DUT, such as nanotube, nanowire, or microresonator, can be attached.
After aligning to the spectrometer, now fabricated on a silicon chip and fixed on the XY table, the DUT is deposited by mechanical friction.


\section{\label{app:coupling}Spectrometer coupling}

The mutual inductance and coupling coefficient between the rf-SQUID and the spectrometer will depend on the circuit geometry, especially the lateral and perpendicular separation between the two superconducting loops containing the Josephson junctions and DUT.
The coupling can be estimated from formulas~\cite{Grover} or calculated numerically~\cite{Khapaev_2015}.
The flip chip assembly setup, \cref{app:flipchip}, allows lateral alignment of both loops with the XY-translation stage.
However because the two chips are attached together the perpendicular ($\Delta Z$) separation cannot be controlled with the flip-chip assembly.
Full XYZ control, in addition to in-situ tuning of coupling, can be achieved in a scanning spectrometer setup.

\begin{figure}
  \centering
  \includegraphics[width=\columnwidth]{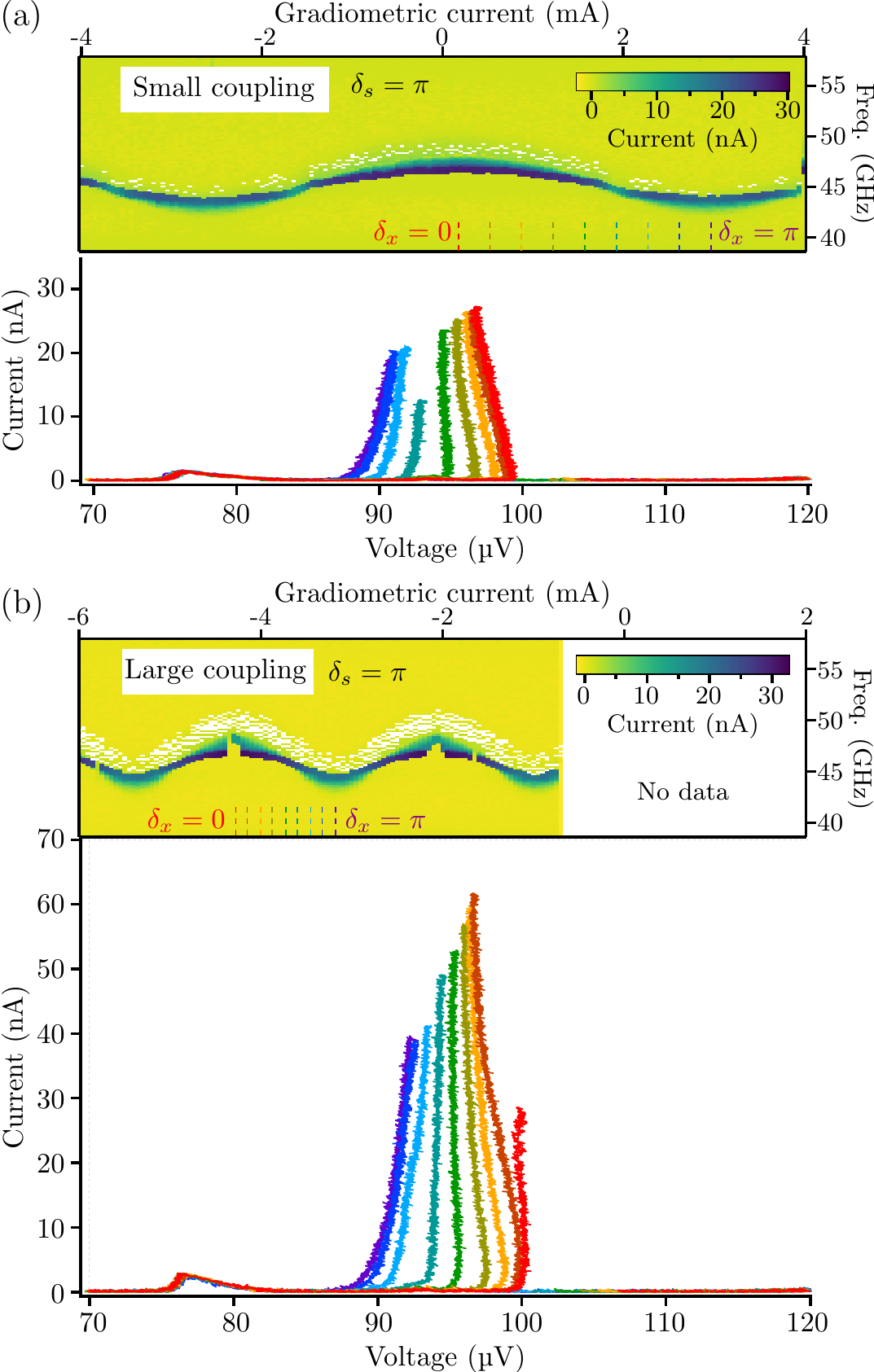}
  \caption{\label{fig:spatial-coupling}Spatial dependence of spectrometer-DUT coupling.
    Absorption peak of rf-SQUID at large (a, ``reduced coupling'') and small separation (b, ``large coupling'') from spectrometer loop, at maximum spectrometer power ($\delta_{s}=\pi$).
    To facilitate comparison, the scaling of the current, voltage, and gradiometric current axes are identical.
    The color coding of the current-voltage characteristics corresponds to the spectrometer phases indicated in the insets.
}
\end{figure}

In~\cref{fig:spatial-coupling} spectra of the rf-SQUID DUT are shown for two coupling strengths.
The DUT chip was detached, shifted laterally away from the spectrometer loop to reduce the coupling, and then reattached to the spectrometer chip.
The measured spectral map (top inset) at maximum spectrometer power ($\delta_{s}=\pi$), as well as spectrometer current-voltage characteristics (bottom inset) at several values of DUT phase $\delta_{x}$ (color-coded with dashed vertical lines), are shown in the ``reduced coupling'' configuration in~\cref{fig:spatial-coupling}(a) and ``large coupling'' configuration in~\cref{fig:spatial-coupling}(b).

At equivalent power, the height of the rf-SQUID absorption peaks in the current-voltage characteristics are larger for larger coupling.
In other words it is possible to perform non-linear spectroscopy at smaller output power, or $\delta_{s}$ closer to zero.
The spectral map in (b) also shows jumps, missing at reduced coupling (a), which are characteristic of hybridization of the rf-SQUID with an electromagnetic mode in the environment.
In addition since the local gradiometric line is aligned inside the spectrometer loop [\cref{fig:implementation}], when the DUT is closer to the spectrometer the periodicity of the spectral lines is smaller.
Data presented in the main text were measured in the ``reduced coupling'' configuration.

\section{\label{app:rfSQUID-spectrum}rf-SQUID spectrum}

\begin{figure}
\includegraphics[width=\columnwidth]{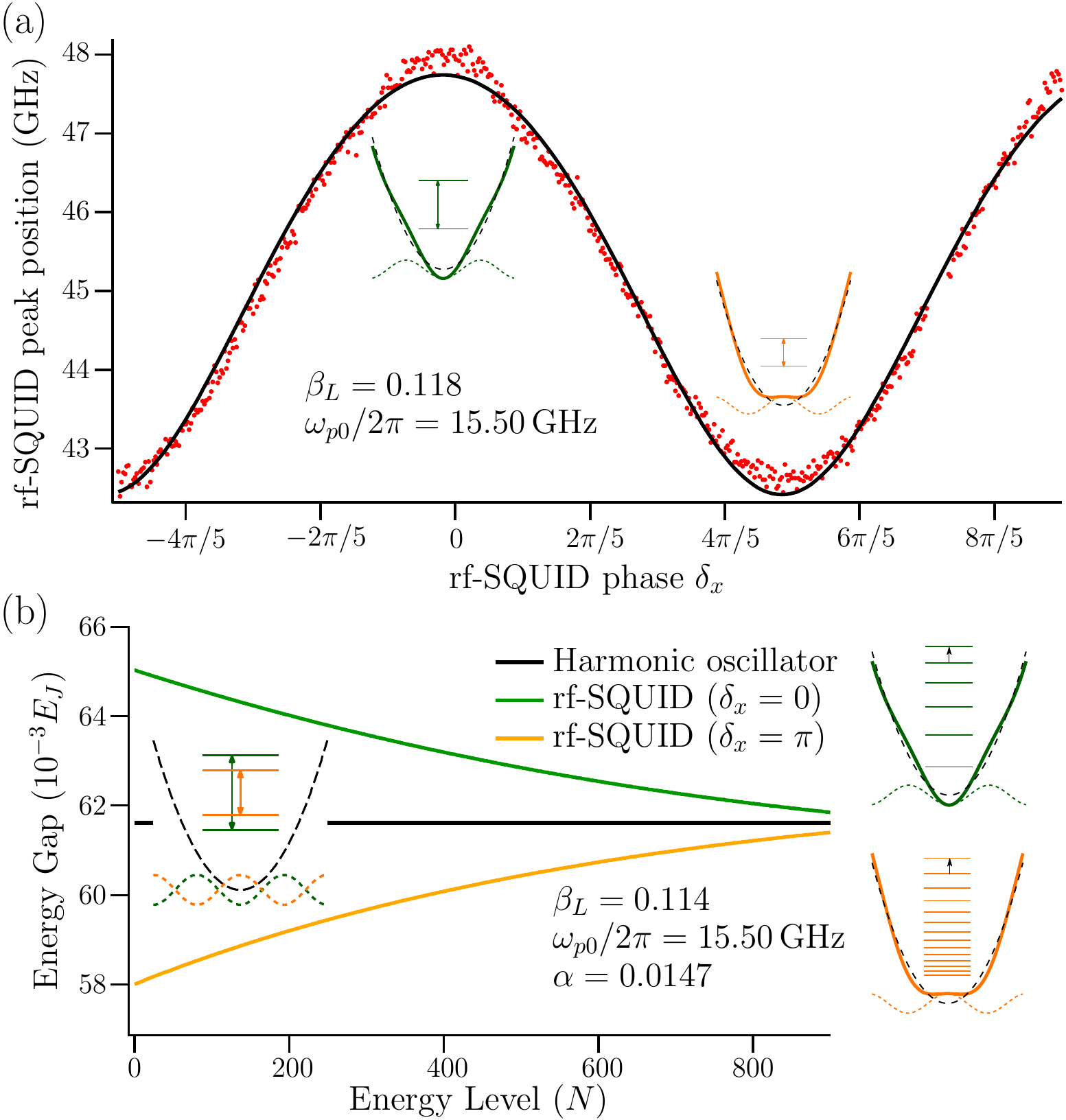}
\caption{\label{fig:spectrum-fit}Energy spectrum of rf-SQUID.
  (a) Frequency of rf-SQUID transition at low spectrometer power ($\delta_{s}=3\pi/10$) extracted from fit shown in~\cref{fig:linear-spectro}(c) (dashed red line).
  (b) Level spacing ($E_{N+1}-E_{N}$) of rf-SQUID as a function of energy level $N$ calculated by diagonalizing the Hamiltonian for two values of phase $\delta_{x} = 0,\pi$.
}
\end{figure}

\cref{fig:spectrum-fit}(a) plots the position of the rf-SQUID absorption peak at low spectrometer power ($\delta_{s}=3\pi/10$) as a function of rf-SQUID phase difference $\delta_{x}$.
The points are obtained by numerically determining the position of the maxima of the peaks in \cref{fig:linear-spectro}(b) and correspond to the red dashed line therein.
The approximately sinusoidal fit to the peak position as a function of $\delta_{x}$ is based on \cref{E1} from the main text.
We rewrite the equation in terms of the plasma frequency, $\omega_{p0}=1/\sqrt{L_{Jx}C_{Jx}}$, which is nominally identical for all of our Josephson junctions,
\begin{equation}
  \label{E2}
  \omega_x(\delta_x)=\omega_{p0}\sqrt{1/\beta_{x}+\cos \delta_x}.
\end{equation}

At $\delta_{x}=0$ the level spacing is larger than for a bare $LC$ mode at frequency $\omega_{x0}=1/\sqrt{L_{x}C_{Jx}}$ due to the increase in the curvature of the potential well (inset level diagram at $\delta_{x}=0$, green solid line) as compared to the bare potential (dashed line).
At $\delta_{x}=\pi$ the curvature is reduced (orange solid line) and the peak position is at its minimum.
The fit for $\beta_L=L_x/L_{Jx}$ is consistent with the $\beta_L = 0.114$ determined from phase calibration, \cref{app:phase-calibration}, and the bare plasma frequency $\omega_{p0}$ is consistent with typical values.

In~\cref{fig:spectrum-fit}(b) the eigenvalues of the Hamiltonian describing the rf-SQUID are calculated numerically with $\omega_{p0}=2\pi\times\SI{15.5}{\GHz}$ and critical current $I_{0x}=\SI{1.5}{\micro\ampere}$, estimated from the spectroscopy measurement and the area of the rf-SQUID Josephson junction.
The plot shows the difference between successive energy levels, $\Delta E=E_{N+1}-E_N$, as a function of level number $N$.
As the spectrometer power increases the average occupation of the rf-SQUID also increases and therefore its absorption frequency changes.
For $\delta_{x}=0$, this frequency will shift to lower values (green line), leading to back-bending [\cref{fig:nonlinear-spectro}(c) and (d)], whereas for $\delta_{x}=\pi$ the frequency increases (orange line), leading to forward-bending [\cref{fig:nonlinear-spectro}(a) and (b)].
As $N$ increases further the level spacing converges to $\hbar\omega_{x0}$, independent of $\delta_{x}$.
The parameter $\alpha=\sqrt{E_c/E_{Jx}}$, where $E_{c}=(2e)^{2}/2C_{Jx}$ is the charging energy, is proportional to the rf-SQUID junction impedance $\sqrt{L_{Jx}/C_{Jx}}$.
A detailed model including the effects of the bias resistor allows fitting the peak bending observed in the spectra~\cite{Dmytruk}.

\section{\label{app:phase-calibration}Phase calibration}
When performing spectroscopy of an inductively coupled device-under-test (DUT) as shown in the schematic \cref{fig:implementation}(b) there are two total magnetic fluxes:  $\Phi_{s}$ for the spectrometer loop and $\Phi_{x}$ for the DUT loop.
The difference in phases of the two Josephson junctions in the spectrometer loop, $\delta_{s}=\delta_{s2}-\delta_{s1}$, is equal to the reduced magnetic flux $\varphi_{s}=2\pi \Phi_{s}/\Phi_{0}$.
In the case of the rf-SQUID DUT, the phase difference $\delta_{x}$ across the Josephson junction in the rf-SQUID loop is the reduced flux $\varphi_{x}=2\pi \Phi_{x}/\Phi_{0}$.
The total flux includes screening by the superconductors and may differ from the applied magnetic flux.
In the data of~\cref{fig:bare-spectro} and~\cref{fig:spectro-meso} screening can be neglected and the applied flux and total flux are interchangeable.

However for superconducting loops such that the geometric loop inductance is larger than the Josephson inductance ($\beta_x=L_x/L_{Jx}\gtrsim1$), in other words big junctions or big loops, the screening fields must be taken into account.
This is the case for spectroscopy of the rf-SQUID (\cref{rfSQUID}).
In the presence of screening we must solve a non-linear matrix equation to obtain the phase differences $\delta_{s}$ and $\delta_{x}$ as a function of the applied magnetic fluxes.

We consider the general case of $N$ control currents used to apply fluxes $\Phi_{An}$ to $N$ loops containing a single Josephson junction of phase difference $\delta_{n}$.
The effective ``coil'' inductances $K_{nm}$ describe the linear relationship between the control currents $I_{m}$ and the flux applied to a given loop $n$:
\begin{equation*}
\Phi_{An} = \sum_m K_{nm} I_m.
\end{equation*}
Each loop has a geometric inductance $L_{n}$, a Josephson inductance $L_{Jn}$, and is coupled to every other loop via a mutual inductance $M_{nl}$.
We include self-screening as well as screening from other loops to obtain the following non-linear equations for the phase differences,
\begin{equation*}
\delta_n = \varphi_{An} - \beta_n \sin \delta_n - \sum_{l\neq n} \gamma_{nl} \beta_l \sin \delta_l,
\end{equation*}
where $\beta_n=L_n/L_{Jn}$, $\gamma_{nl}=M_{nl}/L_n$, and the reduced applied flux is  $\varphi_{An}=2\pi\Phi_{An}/\Phi_{0}$.

In matrix notation we have,
\begin{align}
\label{EA1}\vec{\varphi}_{A} &= \bm{K} 2\pi\vec{I}/\Phi_{0}, \\
\label{EA2}\vec{\delta} &= \vec{\varphi}_{A}-\bm{\gamma}\,\bm{\beta}\sin\vec{\delta},
\end{align}
with matrices,
\begin{align*}
  \bm{K} & =
           \begin{pmatrix}
             K_{11} & K_{12} & \cdots & K_{1N} \\
             K_{21} & K_{22} & \cdots & K_{2N} \\
             \vdots  & \vdots  & \ddots & \vdots  \\
             K_{N1} & K_{N2} & \cdots & K_{NN} 
           \end{pmatrix},
  \\
  \bm{\gamma} &=
           \begin{pmatrix}
             1 & \gamma_{12} & \cdots & \gamma_{1N} \\
             \gamma_{12} & 1 & \cdots & \gamma_{2N} \\
             \vdots  & \vdots  & \ddots & \vdots  \\
             \gamma_{1N} & \gamma_{2N} & \cdots & 1 
           \end{pmatrix},
  \\
  \bm{\beta} &=
           \begin{pmatrix}
             \beta_1 & 0 & \cdots & 0 \\
             0 & \beta_2 & \cdots & 0 \\
             \vdots  & \vdots  & \ddots & \vdots  \\
             0 & 0 & \cdots & \beta_N
           \end{pmatrix}.
\end{align*}

To apply a given configuration of phase differences $\vec{\delta}=(\delta_{1},\ldots,\delta_{N})$ we calculate the necessary applied fluxes $\vec{\varphi}_{A}$ from~\cref{EA2} and then calculate the corresponding control currents $\vec{I}$ by inverting~\cref{EA1}.
As long as the matrix elements $\beta_{n}$ and $\gamma_{nm}$ are sufficiently small there will be a one-to-one correspondence between the coil currents and the junction phase differences.
Our model does not take into account the presence of multiple junctions in a loop, such as for the spectrometer, but this does not affect the results as long as such loops have negligible geometric inductance.

For spectroscopy of the rf-SQUID fluxes are applied via a control current $I_{x}$ flowing through a microfabricated on-chip gradiometric line [\cref{fig:implementation} and \cref{fig:spectro-photo}, orange lines] and a control current flowing through an off-chip superconducting solenoid.
The gradiometric line generates a local magnetic field with a steep field gradient and due to its position along the spectrometer axis predominantly couples magnetic field to the rf-SQUID and not the spectrometer.
The large solenoid creates a spatially uniform magnetic field which couples to both.

\begin{figure}
  \centering
  \includegraphics[width=\columnwidth]{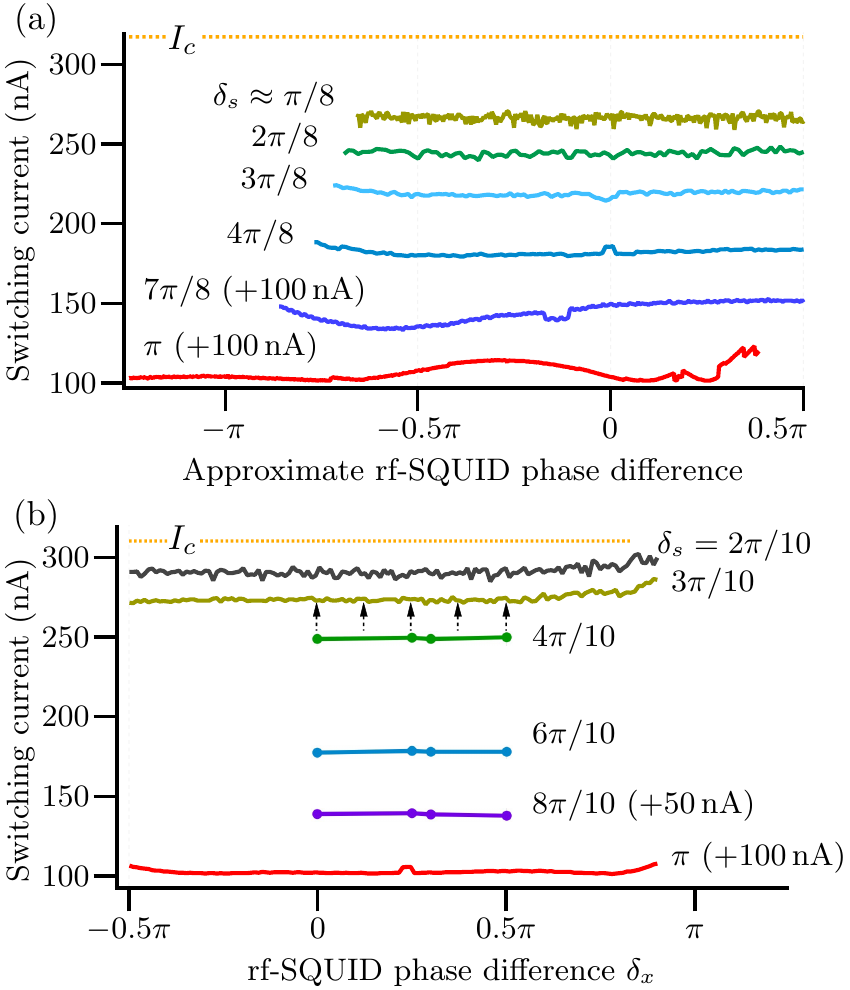}
  \caption{\label{FA5}Effect of phase calibration.
    (a) Correction of linear crosstalk only between the external coil and the gradiometric line.
    Residual modulation of the spectrometer switching current is due to screening by the rf-SQUID.
    (b) The full non-linear inductive interaction between the rf-SQUID and the spectrometer is taken into account.  The calibrated phases $\delta_{s}$ and $\delta_{x}$ are independently controlled [\cref{app:phase-calibration}].}
\end{figure}

To obtain the matrix elements of $\bm{K}$ we tune the control currents iteratively to determine the following surfaces of constant phase difference,
\begin{itemize}
\item $\delta_{s}=\pi$: adjust $\delta_{x}$ through multiples of $2\pi$ while maintaining zero spectrometer supercurrent
\item $\delta_{x}\approx\pi/2$: adjust $\delta_{s}$ through multiples of $2\pi$ while maintaining the rf-SQUID frequency constant and at a point of high sensitivity (large dispersion $d\omega/d\delta_{x}$).
\end{itemize}
The periodicity of sine allows subtracting out the non-linear terms in~\cref{EA2} in order to obtain $\bm{K}$.
Without correction for these terms, equivalent to assuming $\bm{\beta}=0$,  the spectrometer switching current is not flat for ``constant'' $\delta_{s}$, \cref{FA5}(a), indicating that screening by the rf-SQUID loop must be taken into account.

To determine $\bm{\gamma}$ and $\bm{\beta}$ we start with estimated values and iterate until the switching current of the spectrometer does not vary as a function of the expected rf-SQUID phase, \cref{FA5}(b).
We obtain the following matrices,
\begin{equation*}
\bm{K}=\begin{pmatrix}
3450 & 228 \\
1750 & -72 
\end{pmatrix};
\quad
\bm{\gamma}=\begin{pmatrix}
1 & 0.1 \\
0.1 & 1 
\end{pmatrix};
\quad
\bm{\beta}=\begin{pmatrix}
0 & 0 \\
0 & 0.114 
\end{pmatrix},
\end{equation*}
where the units of $\bm{K}$ are $\Phi_0/\si{\ampere}$.

For the measurements shown in~\cref{fig:linear-spectro}(a,b), with $\delta_{s}=3\pi/10$, the supercurrent was maintained at $88\%$ of the maximum value and variations were less than $1\%$ for most of the map.

\section{\label{app:loop-mode}Spectrometer loop mode}

\begin{figure}
\includegraphics[width=\columnwidth]{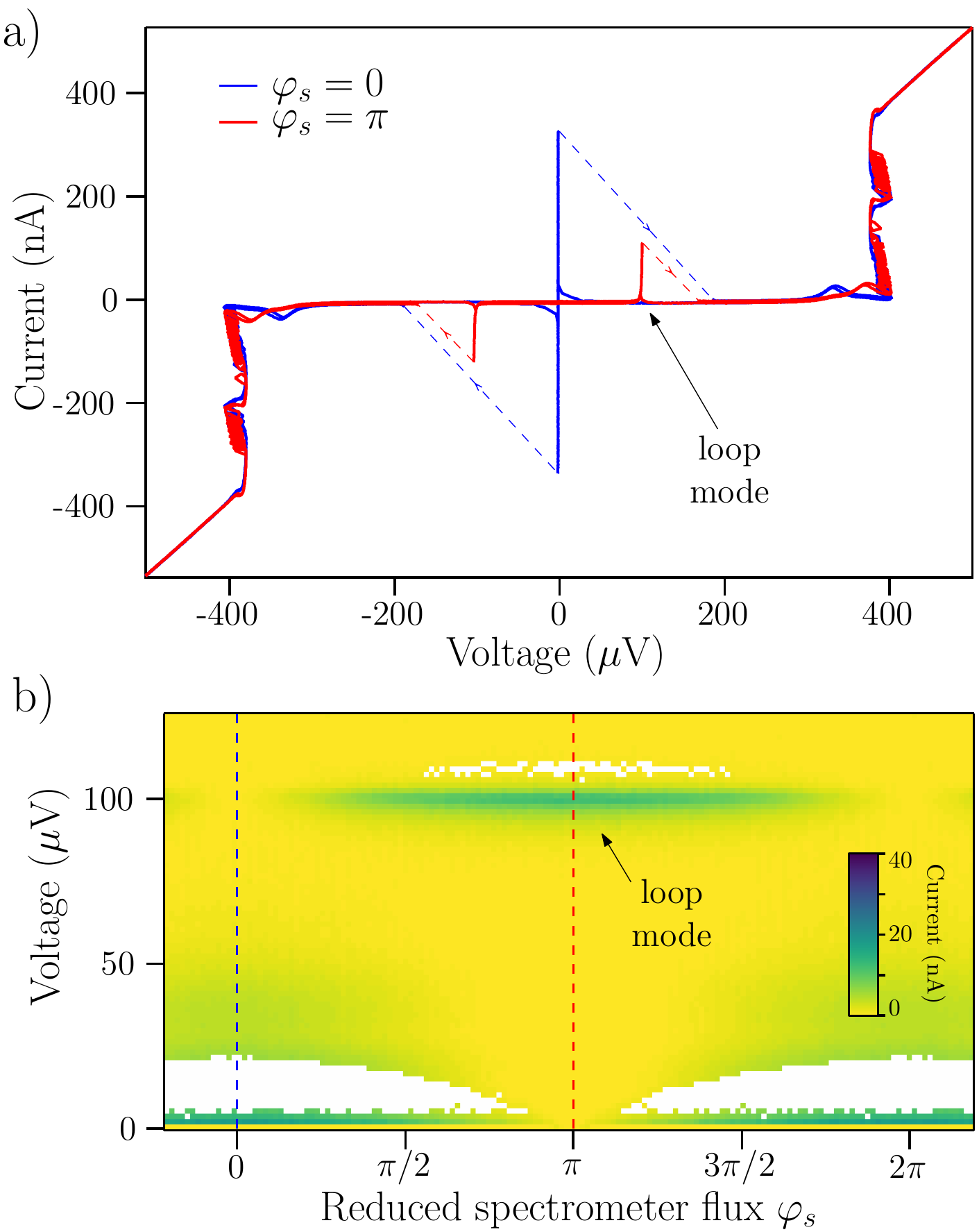}
\caption{\label{fig:loop-mode}
  Spectrometer loop mode.
  (a) Current-voltage characteristic of a spectrometer with visible loop mode resonance at~\SI{100}{\uV} (\SI{48}{\GHz}) for $\varphi_{s}=\pi$.
  The peak is absent for $\varphi_{s}=0$.
  (b) Spectral map shows that loop mode is maximal at $\varphi_{s}=\pi$ and 180° out of phase with the supercurrent peak ($\varphi_{s}=0,2\pi$).
  White regions without data near zero voltage are due to switching from the supercurrent peak.
  The bias is swept toward positive voltages.
}
\end{figure}

The spectrometer circuit has an intrinsic ``loop'' lumped element resonance of angular frequency $1/\sqrt{LC_{J}/2}$ which can only be excited by an oscillating current circulating in the loop~\cite{ZimmermanSullivan,ZappeLandman,SongHurrell,TuckermanMagerlein}.
The loop resonance spectral peak is therefore absent at $\varphi_{s}=0$ and maximal at $\varphi_{s}=\pi$.
This mode is not apparent in~\cref{fig:bare-spectro} because its resonant frequency is too high, exceeding $4\Delta/h\approx\SI{193}{\GHz}$.
By increasing the perimeter of the spectrometer loop or the area of the junctions, thereby increasing $L$ or $C_{J}$ respectively, the resonant frequency may fall in the operating range of the spectrometer, below \SI{200}{\GHz}, and result in a prominent peak, as shown in~\cref{fig:loop-mode}.

The spectrometer whose current-voltage characteristic is plotted in~\cref{fig:loop-mode}(a) has a larger SQUID loop and larger junction capacitances.
The absorption peak due to the loop mode resonance is at~\SI{48}{\GHz} (\SI{100}{\uV}) and is only apparent for $\varphi_{s}=\pi$.
The $\pi$ phase shift of the loop mode peak as compared to the supercurrent peak is highlighted in the spectral map~\cref{fig:loop-mode}(b).

\clearpage
\bibliography{main.bib}

\end{document}